\documentclass[english,aps,pra,reprint,noshowpacs,superscriptaddress]{revtex4-1}   
\usepackage[T1]{fontenc}	
\usepackage[latin9]{inputenc}	
\usepackage{geometry}		
\usepackage{graphicx}
\usepackage[above,below]{placeins}	
\usepackage{times}
\usepackage{hyperref}  
\hypersetup{colorlinks=true,urlcolor=blue,citecolor=blue,linkcolor=blue}   
\begin{document}

\title{Pondering zeros: analysis of a decade of blanks and missed quizzes}
\author{Cassandra Paul}
\affiliation{Department of Physics \& Astronomy, San Jose State University, One Washington Square, San Jose, CA, 95192}
\affiliation{Science Education Program, San Jose State University, One Washington Square, San Jose, CA, 95192}
\author{David J. Webb}
\affiliation{Department of Physics, University of California, Davis, One Shields Avenue, Davis, CA, 95616}
\author{Mary K. Chessey}
\affiliation{Department of Physics, University of California, Davis, One Shields Avenue, Davis, CA, 95616}
\author{James Lucas}
\affiliation{Science Education Program, San Jose State University, One Washington Square, San Jose, CA, 95192}


\begin{abstract}
When assessing student work, graders will often find that some students will leave one or more problems blank on assessments. Since there is no work shown, the grader has no means to evaluate the student's understanding of a particular problem, and thus awards `zero' points. This practice punishes the student behavior of leaving a problem blank, but this zero is not necessarily an accurate assessment of student understanding of a particular topic. While some might argue that this practice is `fair' in that students are aware that they can't receive points for problems they don't attempt, we share evidence that this practice unequally impacts different student groups. We analyze 10 years of UC Davis introductory physics course databases to show that different groups of students (by gender, racial/ethnic group, first generation, etc.) skip problems, and entire exams at different rates. We also share some implications for grading and teaching practices.
\end{abstract}

\maketitle

\section{Introduction}
Grades are used by students, teachers, schools, and much of the rest of society as a measurement of the student's understanding of, or skills in, a subject.  Given this expectation, it's useful to understand the connection between a grade given to a student and the actual understanding or skill of that student. A problem arises when a student fails to provide information about their understanding by skipping a question on a quiz or final exam, or by missing a quiz entirely. Students may leave answers blank for a variety of reasons, which may or may not be related to the student's overall understanding of a topic. When a student leaves a problem blank it may be because they have run out of time, they don't want to reveal their inadequacies,  they don't understand the question,  they don't see value in attempting a problem they don't think they can finish, or for any other number of reasons.  

One way instructors commonly deal with this problem is by awarding the student a zero, reasoning that if the student doesn't provide any evidence of understanding, no credit can be awarded. While on the surface this instructor practice seems logical and even fair, it is still a consequence of the student's  behavior, and not a measurement of their knowledge of a given topic. Thus a zero (the numerical equivalent of no understanding) is awarded for what instead might be better considered missing data. An instructor might be content with awarding a grade that is representative of both classroom behaviors and understanding (in fact the authors of this paper have done this as instructors), however if different student populations engage in blank-leaving behavior at different rates, and this practice negatively impacts some student groups more than others, it could be an important contributor to student performance gaps, and therefore influence student equity in the course.

Various studies \cite{Stephens, Stathopoulou} over the years have investigated the extent to which a mismatch between the cultural norms of a group of students and some factor of the cultural norms of their educational institution may affect learning (and the resulting grades) of that group of students.  For example, Stephens et al. \cite{Stephens} suggest that first generation college students (those who grew up without a college graduate as a parent) are perhaps oriented more toward values necessary for a community and less toward the individual independence that is expected by most institutions of higher education. Taking an exam is perhaps one of the most independent activities a student experiences in a course (especially active learning courses) so this particular cultural difference may be exaggerated when it counts the most, during assessment.

Giving a grade of zero for a blank answer \cite{Reeves} can have a very negative impact on a student's overall exam score, especially if their instructor is using a traditional form of the percent scale \cite{Guskey2013} where earning somewhere around 50 or 60\% of the points is considered failing. In this paper we do not draw any conclusions on why students leave questions unanswered, instead we evaluate the extent to which different student populations engage in this practice. 

In order to evaluate the extent to which student groups are impacted by various classroom practices, a particular model of equity needs to be employed. Rodriguez et al. \cite{Rodriguez2012} describe a model of ``Equity of Fairness'' as one where an intervention has the same impact on one group of students as another. In this paper we apply this model to the practice of assigning zeros for blank solutions and missed quizzes. Should this practice be equitable under this specific model, we postulate that no groups of students would be more likely to engage in this behavior than any other.

For this exploratory study, we chose to consider a few broad groupings of students: males, females, students from race or ethnicity groups underrepresented in STEM, and first-generation college students.
Specifically we ask:

1. How does the behavior of leaving problems blank vary by student population?

2. How does the behavior of missing an exam vary by student population? 

3. Do these behaviors correlate with other measures for understanding?

We examine exam grades in two introductory college physics courses given over ten years and show that different populations of students engage in these behaviors at different rates, and provide evidence that this is not necessarily due to lack of student understanding. Finally, we provide suggestions for instructors concerned with increasing grading equity in their classrooms.


\section{Methods}

\textbf{Data and data sources:} For a separate study \cite{Webb} of differences between grade scales, we looked up 96 original class databases from 2003-2012 to compile a set of 794,088 grades given, on individual parts of 606 quizzes and 76 final exams, to 15,207 students taking Physics 7A and/or Physics 7B \cite{Potter2014} during that set of years.  To these grade data we included student's self-reported demographic data that we received from UCDavis administrative sources. We use this same data set in the current paper.  All averages and standard errors were calculated in Excel and we used STATA software for the statistical tests. We are fortunate that this course utilizes the ``Grading by Response Category'' \cite{Paul2014} method, where student solutions are placed in categories that are later assigned a numeric grade, so that the 96 class-level databases include the individual grades that were given for each problem, the total exam scores for each student, the calculations that led to overall exam scores, and (in many cases) the calculations that led to a final grade.

We limit our results to students who were either US citizens or had permanent resident status. Our administration considers first-generation students to be those for whom neither parent had earned a bachelor's degree or higher. For students from underrepresented minority groups (URM) we use the categorization from our previous work \cite{Paul2017} (which uses information published by the American Physical Society \cite{APSURM}) to indicate those who have ethnic or racial identities underrepresented in physics and STEM; the groups African American, Native American, Latina/o American, Mexican American, Chicana/o, and Pacific Island American are included in this group hereafter referred to as URM.

The CLASP curriculum at UC Davis is spread over three quarters (equivalent to two semesters). We consider only the first two quarters (Physics 7A and 7B) of this 3-quarter introductory physics series because many students were only required to take the first two, and therefore the third quarter (Physics 7C) represents a different sample of students. We examine data from these two courses offered from 2003 through 2012. Throughout these ten years students enrolled in a course spent about 4.7 hours per week in an active-learning discussion/lab section and about 1.3 hours per week in lecture.  The topics covered and the student activities did not change much over these years so the core of this active-learning course did not change.  These courses were usually assigned two instructors for each 300 (or so) students and had a total of 60 different instructors and of order 200-250 Teaching Assistants over these ten years.

\section{Analysis and results}

\begin{figure}
\includegraphics[trim=2.5cm 2.6cm 4.5cm 3.5cm, clip=true,width=\linewidth]{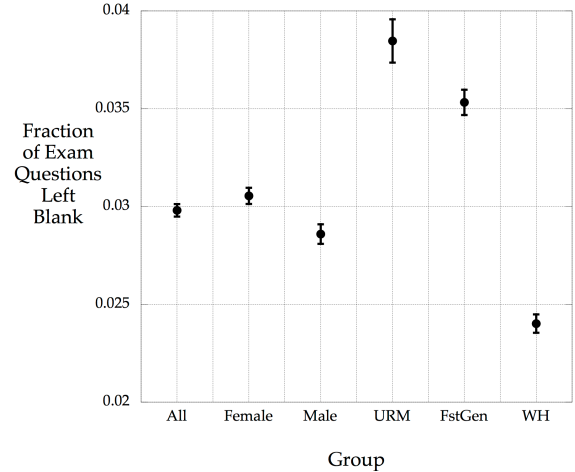}
\caption{The fraction of sub-exam-level answers (on quizzes and finals) that are blank and so were given zeros (error bars are $\pm$ standard errors).  These fractions are plotted for several large groups of students.  URM refers to racial/ethnic groups who are underrepresented in physics and STEM, FstGen are students with neither parent a college graduate, and WH refers to White (non-Hispanic) students. \label{fig1}}
\end{figure}

The number of blank solutions on quiz and final exam items is low overall. The students in these courses left blank only about 3\% of problems on exams that they took. However, when we break the fraction down by student groups we find significant differences. The average fraction of blank answers left by students from racial/ethnic groups who are underrepresented in STEM (URM) is over 1.3 times the average for non-URM and 1.6 times the average for white students.  The average fraction of blank answers left by students who are the first college generation in their family is about 1.3 times the average over non-first-generation. Furthermore, there is a small, but significant ($t=2.97, df=23,430, p=0.003$) difference between male and female blank-leaving behavior; females left more problems blank than males. White students leave the fewest number of blank solutions among the groups we chose to examine.  Note that about 16\% of white students are first-generation and that that percentage is less than half that of any of the other group shown. These results are shown in Fig. \ref{fig1}.

\begin{figure}
\includegraphics[trim=2.5cm 2.6cm 4.5cm 3.5cm, clip=true, width=\linewidth]{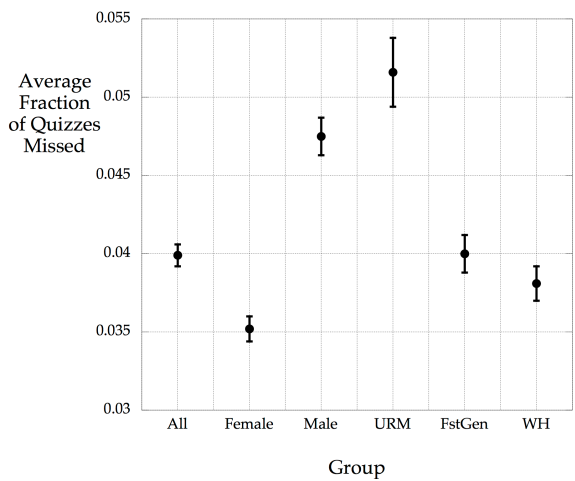}
\caption{The fraction of quizzes missed averaged over each of several large groups of students (error bars are $\pm$ standard errors). Students took from 4 to 9 quizzes, depending upon the instructor, during a course.\label{fig2}}
\end{figure}

The second way that teachers encounter missing data when evaluating students' understanding is when students skip a quiz completely. The fraction of missed quizzes is also low, at about 4\% overall. However, we again see clear differences between different student groups. The average fraction of missed quizzes for students from URM groups is again about 1.3 times the fraction for non-URM students. However, missed-quiz behavior seems to be very different than blank-leaving behavior across the other student groups. For example, while male students left fewer blank solutions than female students left, male students skipped, on average, about 1.3 times as many quizzes as female students skipped. Also, while first-generation students average more blank solutions than the overall average, their average number of missed quizzes is the same as the overall average. These results are displayed in Fig. \ref{fig2}.

\begin{figure}
\includegraphics[trim=2.5cm 2.6cm 4.5cm 3.5cm, clip=true, width=\linewidth]{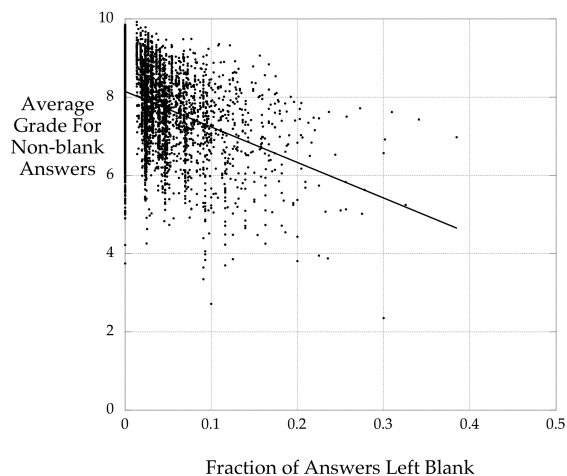}
\caption{For courses graded using a 10-point scale we plot the unweighted average grade for all student answers that were not blank as a function of the fraction of answers that that student left blank.  The line is a linear fit  ($R^2 = 0.13$) showing the trend.\label{fig4}}
\end{figure}

Fig's \ref{fig1} and \ref{fig2} clearly show that different student groups engaged in these two behaviors at different rates but, since achievement gaps between some of these groups are well documented \cite{STEMStats}, one still might question whether leaving a blank for an answer or skipping an exam is not an issue of behavior but of academic strength. In other words, students may leave answers blank because they actually could not have written anything to get them a higher score, or skip exams because they would have received an F anyway.  While it's impossible to make a definite statement about this kind of issue, we can use the dataset to look for correlations between blank-leaving behavior and other student work. When we plot average grade for non-blank-solutions against the overall fraction of answers left blank, we do see that there is a slight negative correlation, meaning that students who leave more blanks are less likely to do well on other course material.  However, an $R^2=0.13$ suggests that the variable ``Fraction of Answers Left Blank'' is a poor predictor of general understanding of the material.  Even the group of students leaving 10 to 15\% of their answers blank includes a broad cross-section (SD = 0.8) of success in completion of the rest of their work, covering all letter grades A through F. See Fig. \ref{fig4}.


\begin{figure}[t]
\includegraphics[trim=2.5cm 2.6cm 4.5cm 3.5cm, clip=true, width=\linewidth]{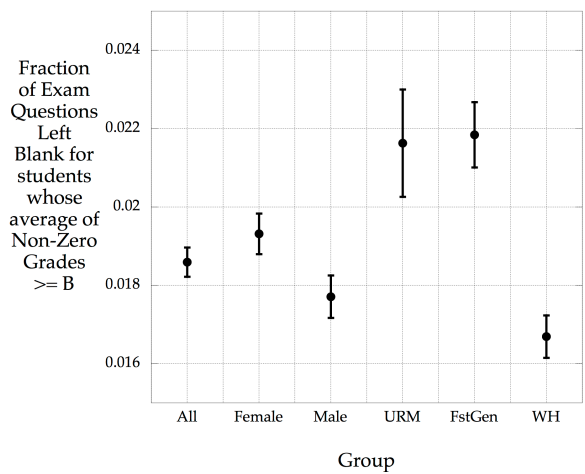}
\caption{For students whose non-zero answers average to a B or better, the fraction of sub-exam-level answers that were given zeros are shown as in Fig. \ref{fig1}.\label{anotherfigure}}
\end{figure}

While the correlation is weak, it still shows that on average, there is at least some relationship between student understanding and blank leaving behaviors. Therefore, we wanted to check to see if blank-leaving was primarily a habit of poor scoring students. For our final analysis of these data, we only consider students who have earned an average of B (85\% with percentage grading or 3.0 on a 4-point scale) or better on all the problems they attempted to solve. (See Fig. \ref{anotherfigure}) While there is less blank-leaving behavior overall, we still find similar differences between the student groups that we saw in Fig. \ref{fig1} which included all students. 

While a complete quantitative analysis of the impact of these zeros on student course grades is beyond the scope of this paper, with the data shown you can begin to interpret consequences of behavioral zeros. For example, in Fig. \ref{fig4}, we see that there are a large number of students who have earned above 85\% (a ``B'') on their submitted work, but have left between 5 and 15\% of their solutions blank. If we assume that each quiz item is weighted the same, we find that these students likely earned a course grade between 72\% (\mbox{``C-''}) and 81\% (``B-'').  Similarly, students who earned a 75\% on submitted work (a ``C'') but left 10\% of their answers blank, likely earned about 68\% (a ``D+'').



\section{Discussion}

For any particular student, leaving a blank or skipping an exam may not change their grade.  In our statement that leaving blanks or skipping a quiz will lead to a lower grade we are simply asserting that neither of these actions will ever raise a grade so that these actions can only lower the grade distribution for the relevant group. We find that some groups of students are more likely to leave blanks than others, and these differences seem to be at least somewhat independent of their overall course knowledge. Future work will use the ``Equity of Fairness'' \cite{Rodriguez2012} model again to examine the extent to which these two behaviors impact students' overall course grades. Student course performance is important, but is not the only reason to consider zeros. There are also other negative consequences we have yet to consider. For example, behavioral zeros can harm student self-concept, self-efficacy and motivation \cite{SelfEff}, which may affect course retention or STEM identity. These relationships are also important to consider when determining grading policies. 

We don't have any direct evidence that the reasons that different groups of students leave blanks or skip exams are different. But the fact that different groups engage in these behaviors at different rates indicates that there are likely equity issues to consider. Furthermore, the fact that (for example) females, the group least likely to skip exams, are not also the student group least likely to skip problems, indicates that these types of behaviors happen for different reasons and perhaps should not be awarded the same zero as a consequence.

To illustrate the importance of thinking about zeros, we share a quote from an upper division undergraduate physics major explaining how they felt about attempting to get partial credit on exams when unsure about solving a problem completely accurately \cite{MaryDiss2018}.

``\textit{Physics is not bullshit. You shouldn't be writing bullshit answers, but you have to. I don't know, it feels a little dirty but at the same time I want a good grade on an exam. \dots Physics is not based on bullshit but here I am.}''

This student needed to sacrifice their value of submitting only high-quality solutions about which they felt confident because they realized that they needed to write something, even what they considered guesswork, to earn credit on the exam. Indeed, we would not want to submit papers for peer-review based on possibly fraudulent assumptions or analysis, so why do we expect this of our students in test-taking scenarios? Furthermore, playing this `dirty' academic game made this student feel like they were succeeding as a physics major in part due to their `bullshit answers,' which created a conflict between physics ideals and classroom experience, and could influence this student's choice to pursue physics in the future.

As instructors we want students to demonstrate the full extent of their understanding on exams so that each student can receive a grade that best represents their knowledge and skill in the subject, but we can also see the issue from the point of view of a student who does not want to pretend to possess knowledge or skill that they feel they do not have. This is only one possible reason students may leave blanks, but it illustrates the need to examine grading student behaviors that are not necessarily representative of their knowledge, especially since blank-leaving behavior may be more common for some groups of students than for others.

Instructors who wish to eliminate the use of behavioral zeros have some options. One possible solution is to adopt standards-based grading \cite{StandBsdGrdng}. While this may be challenging for large lecture courses, it is possible. Other options are minimum grading \cite{Guskey2013}, the practice of never assigning grades lower than 50\% (which is still failing) or using a GPA-like 4.0 scale to minimize the impact of behavioral zeros.

\acknowledgments{We thank the Registrar at UCDavis for providing us with demographic data and the UCDavis and SJSU PER groups for useful feedback on this work.}



\end{document}